\documentclass[12pt]{article}
\usepackage{amsfonts}
\usepackage{amssymb}

\makeatletter

\@addtoreset{equation}{section}
\def\section{\@startsection {section}{1}{\z@}{-2.5ex plus -1ex minus
 -.2ex}{1.3ex plus .2ex}{\large\bf}}
\def\subsection{\@startsection{subsection}{2}{\z@}{-1.5ex plus%
 -1ex minus -.2ex}{1.0ex plus .2ex}{\bf}}

\advance \voffset by -0.9in
\advance \hoffset by -0.6in
\def\rinv{s}
\newcommand{\p}{\partial}
\newcommand{\NN}{\mathbb{N}}
\newcommand{\ZZ}{\mathbb{Z}}

\newcommand{\RR}{\mathbb{R}}

\newcommand{\tr}{{\rm tr}}
\def\bea{\begin{eqnarray}}
\def\eea{\end{eqnarray}}

\begin{document}
\begin{flushright}
EMPG-02-23\\
HWM-02-38\\
hep-th/0212076
\end{flushright}
\vspace{0.5cm}
\begin{center}
\baselineskip 24 pt
{\Large \bf On the existence of minima in the Skyrme model}

\vspace{1cm}
\baselineskip 16pt

{\large B.~J.~Schroers} \\
Department of Mathematics, Heriot-Watt University, \\
Riccarton, Edinburgh EH14 4AS, UK \\
e-mail \,\,{\tt bernd@ma.hw.ac.uk}

\vspace{0.5cm}
{20 November 2002}
\end{center}
\parskip 8pt
\parindent 6pt

\begin{abstract}
\noindent Well-separated Skyrme solitons of arbitrary degree
attract after a  suitable relative rotation in space and
iso-space, provided the orders of the solitons'  leading multipoles
do not differ by more than two.
I summarise  the derivation of this result, obtained jointly with
Manton and Singer,  and
 discuss to what extent its combination
 with earlier results of Esteban  allows one to deduce
the existence
of minima  of the Skyrme energy functional.
\end{abstract}

 \begin{center}
{\small 
Talk given at the  workshop on Integrable Theories, Solitons and Duality}

{\small Sao Paulo, July 2002.}
\end{center}

\vspace{0.1cm}

\section{Introduction}

The Skyrme model is a non-linear field theory  in which
nuclei are modelled by quantum states of classical topological
 solitons \cite{Skyrme}. In this talk I will use the term topological soliton
 for minimal energy solutions  in a field theory which have
an associated integer degree or topological charge. The degree
is conserved for topological reasons and ensures the stability
of the soliton. In the Skyrme model static configurations are maps
 \bea
U: \RR^3 \rightarrow SU(2).
\eea
Points in $\RR^3$ will be denoted by $x=(x_1,x_2,x_3)$
and I  write $r$ for  the Euclidean length
$|x|=\sqrt{x_1^2+x_2^2 +x_3^2}$. Physically, $U$ combines the pion
fields $\pi_1,\pi_2,\pi_3$ and the $\sigma$-field
\bea
\label{pions}
U(x) =   \sigma (x) + i\pi_a(x)\tau_a,
\eea
where $\tau_a$, $a=1,2,3$,  are the Pauli matrices and
 the constraint $\sigma^2 + \pi_1^2+
\pi_2^2+ \pi_3^2 = 1$ is imposed.
Here and in the following, summation over  repeated indices
is always implied. The Skyrme energy functional is
\bea
\label{skyrpot}
E[U] = -\int d^3x \, \left({1\over 2}\tr(L_i L_i) + {1 \over 16}
       \tr(\lbrack L_j,L_i\rbrack \lbrack L_j,L_i\rbrack ) \right),
\eea
where
\bea
\label{Lc}
L_i =  U^{\dagger}\partial_iU
\eea
and  $\partial_i=\partial/\partial x_i$ with $i=1,2,3$.
The Euler-Lagrange equation can be  expressed
in terms of  the modified currents
\bea
\label{LLc}
\tilde{L}_i =  L_i -{1\over 4 }
\lbrack L_j,\lbrack L_j,L_i\rbrack \rbrack
\eea
 and reads
\bea
\label{ELL}
\partial_i \tilde{L}_i =   0.
\eea

It was already noted by Skyrme that the finite-energy requirement
means that the Skyrme fields have to tend to a constant at spatial
infinity, thus becoming effectively maps from $S^3$ to $SU(2)\simeq
S^3$ with  an associated integer degree. Skyrme identified the
degree physically with the baryon number.
 The first
rigorous proof that for finite-energy Skyrme configuration
the degree
\bea
\label{degreeint}
\mbox{deg}[U]=
-\frac{1}{24 \pi^2}\int d^3 x\, \epsilon_{ijk}\,\tr\left(L_iL_jL_k\right)
\eea
is an integer was only given  fairly recently in  \cite{EM}.
As a result    the configuration space
\bea
{\cal C}  =\{ U: \RR^3 \rightarrow SU(2)\,|\, E[U] < \infty \}
\eea
is partitioned into sectors
${\cal C}_k$ consisting of all finite energy configurations
of degree $k$. Faddeev's bound \cite{Faddeev}
\bea
\label{bbelow}
E[U] > 12\pi^2 |k|
\eea
(the strict inequality follows from \cite{Manton})
implies the existence of infima
\bea
\label{infdef}
I_k=\mbox{inf}\{ E[U]\,|\,U\in {\cal C}_k\}.
\eea
The central question of this talk is whether the infima are attained,
i.e. whether minima exist in all sectors of the Skyrme model.
In the following I  call  minimal energy solutions
of non-vanishing degree Skyrme solitons.

To illustrate quite how little is known  rigorously about
the existence of Skyrme solitons
 note that the hedgehog ansatz
\bea
\label{hedge}
U_H(x)=\exp (i f(r)\hat{x}_a\tau_a)
\eea
with the boundary condition  $f(0)=\pi$ and $f(\infty)=0$ leads to
a minimisation problem for the profile function $f$, which
is known to have a unique solution \cite{KL}, called the Skyrmion.
It is also known \cite{Esteban} that the Skyrme energy functional
has a minimum in the sector ${\cal C}_1$.
However, it is still not known whether the minimum is of the
spherically symmetric hedgehog form (\ref{hedge}).

Physically,  we may think of Skyrme solitons of degree $k>1$ as bound
states of Skyrmions. One therefore expects the questions of whether
such bound states exist to be related to the existence of attractive
forces in the Skyrme model. This expectation is borne out by
the analytical work of Esteban in \cite{Esteban}.
Esteban showed that  for a suitable class   functions
\bea
\label{winequality}
I_k\leq I_l+ I_{k-l}
\eea
for all $k,l\in \ZZ$.
Physically this result  is only sufficient to ensure threshold bound states
- it would be satisfied if for example $I_k=C|k|$ for some constant
$C$.  Esteban went on to show that a minimum exist in ${\cal C}_k$
{ \it provided  one assumes} the strict inequality
\bea
\label{inequality}
I_k<I_l+ I_{k-l}
\eea
for all $k\in\ZZ-\{0,\pm 1\}$ and $l\in \ZZ-\{0,k\}$
in the range $|l| + |k-l|< \sqrt 2 |k|$. This result will be
referred to as Esteban's theorem in the following discussion.
In  \cite{EM} it was shown that the result still holds if one
widens the class of allowed functions,
but the inequality (\ref{inequality}) remains a necessary assumption
in the proof. In the cases where the infima $I_l$ and $I_{k-l}$
are attained by Skyrme solitons, the inequality (\ref{inequality})
is equivalent to the existence of attractive forces between those
Skyrme solitons. In \cite{MSS} the existence of attractive
forces is studied in detail
 for Skyrme solitons obeying certain regularity assumptions.
In the second part of this talk I review the arguments
and results of that paper and in the third part
I discuss their  relation  to Esteban's result.

Before plunging into the technical analysis I should  point out that
much is known numerically
about  Skyrme solitons. Numerical searches aided by analytical ans\"atze
suggest  the
existence of Skyrme solitons for
$1\leq k \leq 22$ \cite{BCT,BS1,BS2}. However, it is also worth
stressing that the existence of classical minima does not ensure
the existence of a quantum bound state. This is relevant
in the physical application of the Skyrme model because
nuclei do not exist for arbitrarily large baryon number. The
study of the deuteron as a quantum state of the $k=2$ toroidal
Skyrme soliton in \cite{LMS}  furthermore shows that when
 such a quantum bound state exists the  quantum mechanical
matter distribution
may be quite different from that of the underlying classical soliton.

\section{The interaction energy of  Skyrme solitons}

\subsection{Asymptotics of a Skyrme soliton}
The goal of this subsection is to study the behaviour of
Skyrme solitons near  spatial infinity.  The key step is to think of
the 2-sphere at infinity as a boundary of $\RR^3$ and to show that
the  Skyrme equation is regular there. It then
follows from a unique continuation
argument  that non-trivial
solutions of the Skyrme equation have a non-trivial expansion near
infinity.  In other words,  Skyrme solitons necessarily
 have a non-trivial large $r$ expansion in powers
of $1/r$, possibly combined with $\ln r$.

To explain the basic ideas,
consider a finite point $x_0$ and assume (if necessary redefining
$U(x) \rightarrow U(x_0)^{-1}U(x)$ ) that $U(x_0)=1$. Then we have
the expansion
\bea
U(x) = 1 +u(x)
\eea
in a neighbourhood of $x_0$,
with the $2\times 2$ complex matrix $u$
satisfying  the algebraic constraints
\bea
\label{littlu}
u + u^\dag + uu^\dag = 0, \quad \tr(u) +\mbox{det} (u)=0 .
\eea
After lengthy algebra, the Skyrme equation can be rewritten
\bea
\label{rewritten}
P(u,\p u, \p^2 u)= Q(u,\p u) + F(u,\p u),
\eea
where $P$ is linear in the second derivatives $\partial^2 u$ and contains
the Laplace operator $\Delta u$. The functions $Q$ and $F$  crucially do not
depend on $\partial^2 u$ and are, respectively,  quadratic and
of degree four in $u$. The point of  writing the Skyrme equation in
this way is that the operator
\bea
f \mapsto P(u,\partial u,\partial^2 f)
\eea
is linear and elliptic near $x_0$. If we assume that both $u$ and the currents
$L_i$ are H\"older continuous,  we deduce by elliptic regularity
that a solution of (\ref{rewritten}) is  twice differentiable.
Bootstrapping further, one concludes   that any solution is in fact
smooth (i.e. $C^{\infty}$). The key idea in this procedure is to
re-interpret the non-linear Skyrme equation as an elliptic linear
equation with coefficients depending on $u$ and $\partial u$.

In order to apply the same idea at spatial infinity, we treat
the 2-sphere at spatial infinity as a boundary of $\RR^3$ and introduce
coordinates $(s,\theta, \varphi)$, where $s=1/r$ and $\theta$ and $\varphi$
are the usual spherical coordinates. Then defining
\bea
D_i = r\partial_i = \frac{1}{\rinv} \partial_i
\eea
the Euclidean Laplacian takes the form
\bea
\label{Lapla}
\Delta  = \rinv^2\Delta_b
\eea
with
\bea
\Delta_b = \rinv^2 \p_\rinv^2  +
\Delta_\omega,
\eea
where $\Delta_{\omega}$ is the Laplacian of the unit 2-sphere.
Now we write $U = 1 +u$ for large $r$ (that is, for small positive
$\rinv$)
and obtain a '$b$'-version of the Skyrme equation
\bea
\label{inftyeq}
P_b(u,D u,D^2u) = Q_b(u,D u) + \rinv^2 F_b(u,D u).
\eea
This equation can now be analysed using tools from the theory
of so-called $b$-differential operators, see \cite{Melrose,Mazzeo}.
 In particular one
can use the boundary conditions that $U$ tends to one at $\infty$
and that the currents $L_i$ vanish there to solve (\ref{inftyeq})
iteratively, thus producing an expansion of the solution in powers
of $s=1/r$, with possible factors $\ln r$ in higher order terms.
At this stage  we have an asymptotic expansion of Skyrme solitons,
but do not yet know whether there is a leading non-zero term in
the expansion. This can be established with the help of a unique
continuation theorem \cite{Hormander}, suitably
adapted for our purposes.

The upshot of this chain of arguments, described in detail
in \cite{MSS}, is   that every  Skyrme
soliton has a leading
Lie-algebra valued multipole field
\bea
\label{genmultipole}
u_M(x) =i\tau_a\sum_{m=-M}^M\frac{4\pi}{2M+1} Q^a_{Mm}\frac{Y_{Mm}
(\theta,\varphi)}{r^{M+1}},
\eea
where $Y_{Mm}$ are the usual spherical harmonics on $S^2$. In accordance
with the usual nomenclature we refer to the multipole field
(\ref{genmultipole})
as a $2^M$-pole and call $M$ the order of the multipole.
The leading multipole  moments $Q^a_{Mm}$ are  independent of the location of
the Skyrme soliton,  and  are acted on naturally by rotations
and iso-rotations.  They are  crucial for the calculations in the
following section. Note that it was already shown in \cite{Mantonn}
that the leading multipole cannot be a monopole. The Skyrmion field
({\ref{hedge}) is known to have an iso-triplet of dipoles as
leading multipoles. For other Skyrme solitons, too, the leading
multipoles have been investigated to some extent. The leading
multipole of the largest  known order  is an octupole (i.e. multipole
order 3)  which
is believed to arise  in the charge  seven
icosahedral Skyrme soliton \cite{BS1}

\subsection{Interaction energy of two scalar multipoles}
As  a technical preparation for the computation of the interaction
energy of two Skyrme solitons we derive a formula for the interaction
energy of two scalar multipoles. In our description of multipoles
we use the conventions of \cite{Jackson} throughout.
Consider the field of a $2^M$-pole  centred at $X_+=(0,0,R/2)$,
where $R>0$. In terms of spherical coordinates $(\theta_+,\varphi_+)$
centred at $X_+$ it reads
\bea
f_M(x)=\frac{4\pi }{(2M+1)}
\sum_{m=-M}^{M}Q_{Mm}\frac{Y_{Mm}(\theta_+,\varphi_+)}{|x-X_+|^{M+1}}.
\eea
Similarly define  $X_-=(0,0,-R/2)$ and let  $(\theta_-,\varphi_-)$ be
spherical coordinates centred at $X_-$. A $2^N$-pole field
centred at $X_-=(0,0,-R/2)$ has the form
\bea
g_N(x)=\frac{4\pi }{(2N+1)}
\sum_{n=-N}^{N}\tilde Q_{Nn}\frac{Y_{Nn}(\theta_-,\varphi_-)}{|x-X_-|^{N+1}}.
\eea
A natural measure for the interaction energy between harmonic function in
the upper and those in the lower half plane is
\bea
\label{interact}
V[f,g] = \int_{x_3=0} dx_1dx_2\,\,(  g\partial_3 f -f\partial_3 g ).
\eea
The computation of this interaction energy for the multipole fields
$f_M$ and $g_N$ is not easy because the combined  field  of
 the two multipoles only has cylindrical symmetry
about the $x_3$-axis and not the full rotational symmetry of each of
the  multipoles. As explained in \cite{MSS}, Fourier transformation
in the $x_1x_2$-plane turns out to be an efficient tool
for evaluating the integral in (\ref{interact}).
Assuming without loss of generality that $M\leq N$, the answer is
\bea
\label{interformula}
V[f_M,g_N]=
\frac{(4\pi)^2}{R^{M+N+1}}&&\frac
{(M+N)!(-1)^{N+M}}{ \sqrt{(2M+1)(2N+1)} } \,\,\,\times
\nonumber \\ &&
 \sum_{m=-M}^{M}\frac{\bar Q_{Mm} \tilde Q_{Nm}}
 {\sqrt{(M-m)!(M+m)!} \sqrt{(N-m)!(N+m)!}}.
\eea

For us, the most important feature of the formula (\ref{interformula})
is that,  for non-vanishing multipole moments,
the interaction energy can always be made non-zero by a
suitable rotation  of the  multipole of the highest order, in our
case $N$ (if $M=N$ it does not matter which of the multipoles gets
rotated). By definition the multipole
 components $\tilde Q_{Nn}$ are the components of an element  $\tilde Q$ of
the $(2N+1)$-dimensional irreducible
unitary representation $W_N$ of $SO(3)$, with $G\in SO(3)$ acting via
$\tilde Q_{Nn} \mapsto
\sum_{n'=-N}^N U^N_{nn'}(G)\tilde Q_{Nn'}$.
  Think of the pairing of the multipole moments in (\ref{interformula})
  as a
linear form
\bea
\label{multipolemap}
 F_{Q}: W_N \rightarrow \RR, \qquad
 \tilde Q \mapsto \sum_{m=-M}^{M} \frac{\bar Q_{Mm} \tilde Q_{Nm}}
{\sqrt{(M-m)!(M+m)!} \sqrt{(N-m)!(N+m)!}}.
\eea
Then it follows from the irreducibility of $W_N$ that $U^N(G)\tilde Q$ cannot
lie in the kernel of $F_Q$ for all $G$. Therefore
$F_Q(U^N(G)\tilde Q)\neq 0$  for some $G$. Hence
 the interaction energy (\ref{interformula})
is non-vanishing after  rotating $\tilde Q$ with that  $G$.

\subsection{Interaction energy of well-separated Skyrme solitons}
Consider now two  Skyrme solitons  $U^{(1)}$ and
$U^{(2)}$  of  degrees  $k$
and $l$. Since the total energies  of
both $U^{(1)}$ and $U^{(2)}$ are finite there must
be balls  $B_1$ and $B_2$ in $\RR^3$ so that most of the energy
 of $U^{(1)}$ and $U^{(2)}$  is concentrated in,
respectively, $B_1$ and $B_2$. Outside the balls $B_1$ and $B_2$
the asymptotic analysis of the previous section applies. Suppose
that the leading multipole of $U^{(1)}$ is a $2^M$-pole and the
leading multipole of $U^{(2)}$ is a $2^N$-pole. Denoting the
radii of $B_1$ and $B_2$ by $D_1$ and $D_2$ we have
\bea
U^{(1)}(x) \sim 1+u_M(x)\quad\mbox{for}\quad  x\not \in B_1
\eea
and
\bea
U^{(2)}(x) \sim 1+v_N(x)\quad\mbox{for}\quad x\not \in B_2,
\eea
where $u_M$ is of the form (\ref{genmultipole}) and analogously
\bea
\label{genmultipolee}
v_N(x) =i\tau_a\sum_{n=-N}^N\frac{4\pi}{2N+1} \tilde Q^a_{Nn}\frac{Y_{Nn}
(\theta,\varphi)}{r^{N+1}}.
\eea
Using
the translational invariance of the Skyrme energy functional we can
assume  without loss of generality that $B_1$  is centred at
$X_+=(0,0,R/2)$  and   that $B_2$ is centred at $X_-=(0,0,-R/2)$, where $R$ is
so large that $B_1$ and $B_2$  do not overlap, i.e. $R>D_1+D_2$.
The  parameter $R$  will be interpreted  as the separation of the Skyrme
solitons.
Then we define the following product configuration
\bea
\label{product}
U_R(x)=U^{(1)}(x)U^{(2)}(x).
\eea
This configuration  has degree $k+l$ and finite energy,
so that  $U_R \in {\cal C}_{k+l}$. A lengthy calculation
performed in \cite{MSS} shows that
\bea
\label{inenn}
E[U_R]=E[U^{(1)}] +E[U^{(2)}] + \Delta E + {\cal O}\left(\frac {1}
  {R^{2N+4}}\right) +
{\cal O}\left(\frac {1} {R^{2M+4}}\right),
\eea
where
\bea
\label{key}
\Delta E  = 2\sum_{a=1}^3 \int_{x_3=0}dx_1 dx_2 \,
(u_M^a\partial_3v_N^a-v_N^a\partial_3u_M^a).
\eea
Now we note that $\Delta E$ is the sum over iso-components
of the the scalar interaction terms we studied above
 \bea
\Delta E_a = 2\int_{x_3=0} dx_1dx_2 \,\,
( u^a_M \partial_3 v^a_N - v^a_N\partial_3 u^a_M) = -2 V[u^a_M,v^a_N].
\eea
Picking one of the iso-indices, say $a=1$,   we can use iso-rotations
to make sure that  the first iso-components $u^1_M$ and $ v^1_N$
are non-vanishing. The result of the previous subsection
then implies  that  we can make
the multipole interaction energy  $\Delta E_1$
non-zero by a spatial rotation  of  the Skyrme soliton  with the higher
multipole order.

Now consider  the sum
\bea
\Delta E = \Delta E_1+ \Delta E_2 + \Delta E_3.
\eea
Following an idea in \cite{CK}
we would like to show that we can always arrange for  $\Delta E$
to be negative by a suitable iso-rotation of one of the Skyrme
 solitons.
We may assume  that, possibly after re-labelling the pion fields,
\bea
 \Delta E_1\geq \Delta E_2 \geq \Delta E_3.
\eea
If $\Delta E< 0$ we are done, so suppose that $\Delta E \geq 0$.
Since we know that not all $\Delta E_a$ vanish we
can conclude that $\Delta E_1  > 0$.
Now perform an iso-rotation
of Skyrme soliton  $2$  by $180$ degrees  around the third iso-spin axis.
This reverses the sign  of $v_N^1$ and $v_N^2$ and hence  of
$\Delta E_1$  and $\Delta E_2$. The new value of $\Delta E$
is
\bea
\Delta E& =&- \Delta E_1 -\Delta E_2 + \Delta E_3 \nonumber \\
 &=& -\Delta E_1 -(\Delta E_2 - \Delta E_3) < 0,
\eea
since $-\Delta E_1 <0$ and, with our ordering,
$-(\Delta E_2 - \Delta E_3)\leq 0$.

Thus, the contribution $\Delta E$ to the interaction energy
of two Skyrme solitons $U^{(1)}$ and $U^{(2)}$
can always be made  negative
by suitable relative rotations and iso-rotations of the Skyrme solitons.
Returning to the expression  (\ref{inenn}) for the interaction energy
and noting from (\ref{interformula})  that $\Delta E$ falls off like
$R^{-(N+M+1)}$,
we conclude that $\Delta E$ is the leading contribution for large
separation $R$ provided the orders of the leading multipoles do not differ
by more than 2, i.e. provided that
$|N-M|\leq 2$. In that case we can therefore
always arrange for the interaction energy to be negative for sufficiently
large separation. In symbols, for sufficiently
large  $R$ and the appropriate orientations we have
\bea
\label{tops}
E[U_R] < E[U^{(1)}]+ E[U^{(1)}].
\eea

\section{The existence of Skyrme solitons}

The results of the previous section are not sufficient to
 derive the existence minima  in a general sector ${\cal C}_k$
 of the Skyrme model. Nonetheless it is instructive  to see
 how far one can get with the following two extra assumptions.
 \begin{enumerate}
 \item When minima exist they satisfy the technical assumptions made
 in \cite{MSS}. In  particular the field  $U$ and the currents $L_i$
  are H\"older continuous.
  \item The leading multipole of any Skyrme soliton is at most an octupole
  (order 3).
\end{enumerate}
The second assumption seems particularly restrictive and deserves a
comment.
Since monopoles (order 0) cannot arise
in Skyrme solitons and dipoles  (order 1)
are known to arise in some, the restriction to leading multipoles of order
at most 3 means that any two Skyrme solitons we consider in the following
have multipoles whose orders do not differ by more than 2. As mentioned
earlier, none of the numerically studied Skyrme soliton  violate
 assumption 2.

Recall from section 1 that
 Skyrme solitons have rigorously been shown to exist
in  the topological sectors ${\cal C}_1$ and ${\cal C}_{-1}$.
Although it is not clear whether the minimum is of the hedgehog
form (\ref{hedge}), we can use the numerically computed energy
of the Skyrmion to give an upper bound on the infimum $I_1$.
To avoid the discussion of numerical accuracy we use the energy
of the instanton generated hedgehog field \cite{AM} $\tilde I_1
=1.24 \times  12 \pi^2$ to estimate
\bea
I_1 < \frac{5}{4} \,12 \pi^2.
\eea
Using Esteban's weak inequality (\ref{winequality}) we deduce
\bea
I_{k} < \frac{5}{4}\, 12 \pi^2 |k|
\eea
for all $k\in \ZZ.$
Combining this result with the energy bound (\ref{bbelow})
 we also deduce  that
\bea
\label{multiin}
I_{k} < \sum_s I_{l_s}
\eea
for all $ k\in \ZZ$ and $l_s\in \ZZ$, $s\in \NN$ satisfying
$\sum_s l_s =k$ and
$ \sum_s |l_s| \geq \frac{5}{4}|k|$.
Thus we can weaken the assumption in Esteban's theorem about the
existence of minima. The existence of a minimum in sector ${\cal C}_k$
follows if the strict inequality (\ref{inequality}) is satisfied for
all $k\in \ZZ$ and all $l \in \ZZ-\{k,0\}$ satisfying
\bea
\label{newcon}
|l| + |k-l| < \frac{5}{4}|k|.
\eea
Combining this result with our
proof of attraction between Skyrme solitons and the assumptions
made above we can prove
the existence of  minima in the  sectors ${\cal C}_k$ for low values
of $k$.
 We prove the claim by induction
and first consider positive $k$.
 We know from \cite{Esteban} that the minimum exists for
$k=1$. Suppose that minima exist in ${\cal C}_l$ with $l=1,2, ..., k-1$.
Then apply the  product ansatz $U_R$ and the inequality
 (\ref{tops}) to pairs of Skyrme solitons
$U^{(1)}$ and $U^{(2)}$ of degree $l$ and $k-l$ for $l=1, ... ,k-1$.
By assumption the energies of these Skyrme solitons are
equal to the infima
$I_l$ and $I_{k-l}$.
Then, by definition of the infimum $I_k$ (\ref{infdef}) and
from (\ref{tops})
\bea
I_{k}\, \leq \,\,E[U_R]\, <\, E[U^{(1)}]+ E[U^{(1)}] = I_l +I_{(k-l)}
\eea
for all $l=1, ..., k-1$.
As long as $k\leq 8$ the only integers $l$ satisfying the inequality
(\ref{newcon}) are in the range $l=1, ... ,k-1$,
 so that we have satisfied the condition
of Esteban's theorem and can conclude that  the infimum in ${\cal C}_k$
is attained. The claim for negative $k$ follows by applying the
energy preserving reflection map $U(x)\mapsto U^\dagger(-x)$.

For $k>8$ the step in the inductive proof fails because we then
need to consider the inequality (\ref{inequality}) for negative
$l$, too. For example for $k=9$, the condition (\ref{newcon}) is
satisfied for $l=-1$ and $k-l=10$. However, since we do not know
whether Skyrme  solitons exist in  the sector ${\cal C}_{10}$ we
cannot establish the inequality $I_9 <I_{-1}+I_{10}$ with our method.
Numerical evidence and physical intuition suggests that
  the inequality (\ref{inequality}) should certainly hold
when $k$ is positive and $l$ or $(k-l)$ is negative.
If it was violated in that
range it would be energetically favourable
for matter (solitons of positive degree)
and antimatter (solitons of negative degree)
to coexist  in some  sectors  of  the Skyrme model.
However, while this seems unlikely, we cannot disprove the possibility
with our methods. If one could rule it out by other methods, the
inductive argument given above could be extended to
 prove the existence of minima in all sectors of the Skyrme model.

To end, I  point out that it seems plausible   that our results can also
be used to give a more direct proof of the existence of  minima in all
sectors of the  Skyrme model.
The idea of such a proof  is that  the infimum in
one of the sectors, say ${\cal C}_k$,    can only  fail to be attained
if it is energetically favourable  for configurations in that sector
to split into  well-separated Skyrme solitons.
 The infimum $I_k$ would  then
be realised  by a  virtual solution, made up  of infinitely separated Skyrme
solitons. The number of Skyrme solitons
in this virtual solution is necessarily finite. For suppose the
degrees  of the Skyrme solitons are $l_s$, $s\in \NN$  with $\sum_s l_s =k$.
By the inequality (\ref{multiin}) we can only have the equality
\bea
I_k=\sum_s I_{l_s}
\eea
if $\sum_s |l_s| <\frac{5}{4}|k|$, which means that all but a finite
number of the $l_s$ must be zero. Suppose that $l_s\neq 0$ for $s=1,2 ... S$
and $l_s =  0$ for $s>S$.
Physically, one expects the following dichotomy to hold.
\begin{enumerate}
\item[(a)]
For given $k$ the infimum $I_k$ is attained by a
Skyrme soliton of  degree  $k$.
\item[(b)] The infimum $I_k$  is equal
to a  finite sum
\bea
\label{theend}
I_k=\sum_{s=1}^S  E_{l_s}
\eea
of energies $E_{l_s}=I_{l_s}$
 of $S$ Skyrme solitons of degrees $l_s \in \ZZ-\{0,k\}$.
\end{enumerate}
With our assumption on the leading multipoles
our earlier proof of attractive forces in the Skyrme model
implies that we can always lower the energy of a finite number of
infinitely separated Skyrme solitons
by chosing appropriate  orientations
for two of the Skyrme solitons and bringing them closer together.
Thus the infimum $I_k$ cannot be of the form (\ref{theend})
and  if the dichotomy is valid, option (a) must hold.

\section{Conclusion}
In this talk I reviewed the result, obtained jointly with
Manton and Singer, that
arbitrary Skyrme solitons attract for suitable
relative orientation and iso-orientation,   provided the
orders of their leading multipole moments do not differ by more
than two. I also indicated how  this result,
combined with the work
of Esteban, can be used  to shed light on the question of
minima  in the Skyrme model. However, the approaches  sketched
here also show  how difficult it is to establish rigorous results
on minima. In particular the assumption on the multipoles of Skyrme
solitons severely restricts the generality of the sort of argument
outlined above.

\vspace{1cm}

\noindent {\bf Acknowledgments} \,\,\,\
I thank Nick Manton and Michael Singer for collaboration on the
related paper \cite{MSS} and discussions.
 I acknowledge  financial support trough an EPSRC
 advanced research fellowship.

\end{document}